\documentclass[12pt,peerreview,onecolumn]{IEEEtran}

\IEEEoverridecommandlockouts

\usepackage{color}
\usepackage{graphicx}
\usepackage{amsmath}
\usepackage{amssymb}
\usepackage{algorithm}
\usepackage{algorithmic}
\usepackage{amsmath}
\usepackage{multirow}
\usepackage{booktabs}
\usepackage{array}
\usepackage{amsthm}
\usepackage{stfloats}
\usepackage{caption}
\usepackage{bm}
\usepackage{subfigure}

\newcommand{\be}{\begin{equation}}
\newcommand{\ee}{\end{equation}}
\newcommand{\bea}{\begin{eqnarray}}
\newcommand{\eea}{\end{eqnarray}}
\newcommand{\ba}{\begin{array}}
\newcommand{\ea}{\end{array}}

\makeatletter

\newcommand{\Rmnum}[1]{\expandafter\@slowromancap\romannumeral #1@}
\makeatother

\title{Practical Modeling and Beamforming for Intelligent Reflecting Surface Aided Wideband Systems
\thanks{W. Cai, H. Li, and M. Li are with the School of Information and Communication Engineering, Dalian University of Technology, Dalian 116024, China, (e-mail: wenhaocai@mail.dlut.edu.cn, hongyuli@mail.dlut.edu.cn, mli@dlut.edu.cn).}
\thanks{Q. Liu is with the School of Computer Science and Technology, Dalian University of Technology, Dalian 116024, China (e-mail: qianliu@dlut.edu.cn).}
}

\author{Wenhao Cai,
        Hongyu Li,~\IEEEmembership{Student Member,~IEEE,}
        Ming Li,~\IEEEmembership{Senior Member,~IEEE,}
        and Qian Liu,~\IEEEmembership{Member,~IEEE}}

\begin{document}
\maketitle
\thispagestyle{empty}
\begin{abstract}
Intelligent reflecting surface (IRS) has emerged as a revolutionizing solution to enhance wireless communications by intelligently changing the propagation environment. Prior studies on IRS are based on an ideal reflection model with a constant amplitude and a variable phase shift. However, it is difficult and unrealistic to implement an IRS satisfying such ideal reflection model in practical applications. In this letter, we aim to investigate the phase-amplitude-frequency relationship of the reflected signals and propose a practical model of reflection coefficient for an IRS-aided wideband system. Then, based on this practical model, joint transmit power allocation of each subcarrier and IRS beamforming optimization are investigated for an IRS-aided wideband orthogonal frequency-division multiplexing (OFDM) system. Simulation results illustrate the importance of the practical model on the IRS designs and validate the effectiveness of our proposed model.
\end{abstract}


\maketitle
\section{Introduction}
Intelligent reflecting surface (IRS) is deemed as a promising and revolutionizing technology for future wireless communication systems.
As a kind of impedance metasurface, each element of IRS is composed of configurable electromagnetic (EM) internals and
can reflect the incident EM wave with a desired phase shift \cite{Prcatical_IRS}.
Thus, IRS is able to intelligently change the propagation environment and significantly enhance quality and coverage of wireless communications \cite{IEEE 1}-\cite{Online 3}.


The applications of IRS in different wireless communication scenarios have been extensively investigated with different performance metrics, e.g., rate maximization \cite{HY1}, energy efficiency maximization \cite{HY4}, secrecy rate maximization \cite{MY1}, transmit power minimization \cite{HY2}, \cite{MY2}, sum-rate maximization \cite{Hongyu} and weighted sum-rate maximization \cite{NEW}.
In those studies, each IRS element is assumed to have an ideal reflection model, which has a constant amplitude and a variable phase shift. However, the actual reflection coefficient heavily depends on resonance frequency, signal frequency, loss resistance, and the quality factor of the resonance circuit, etc. It is difficult and unrealistic to implement an ideal IRS satisfying such ideal model. Although the ideal model is mathematically simple, it cannot precisely describe the realistic signal reflection by a practical IRS. Consequently, the IRS beamforming design with such inaccurate ideal model will cause severe performance degradation in realistic systems. Therefore, developing a practical model of IRS reflection is very crucial to provide more accurate guidance for the IRS beamforming optimization and other designs.

In \cite{Prcatical_PSM}, a practical phase shift model was introduced to characterize the fundamental relationship between the reflection amplitude and phase shift. Compared to the conventional ideal model, the substantial performance improvement can be achieved by applying the new practical model into an IRS-aided MISO system. Nevertheless, this pioneering work only investigated the amplitude-phase relationship for a narrowband signal. Actually, the power consumption and the phase shift of the incident signal are also associated with the difference between the signal frequency of the incident wave and the resonant frequency of the reflecting circuit. When a wideband signal is reflected by IRS elements, the reflection of EM waves will cause dispersion, just like white light passing through a prism. Thus, signals with different frequencies exhibit reflection coefficients with different amplitudes and phase shifts.


In this letter, we investigate the phase-amplitude-frequency relationship of the reflected signals and propose a practical model of reflection coefficient for a wideband IRS system. Then, in order to evaluate the importance and effectiveness of the practical model, the system performance using this proposed practical model for IRS beamforming design is studied.
Considering an IRS-aided wideband frequency-division multiplexing (OFDM) system, we aim to maximize its average achievable rate by jointly optimizing the transmit power allocation of each subcarrier and the IRS beamforming. An iterative algorithm is introduced to sub-optimally solve this non-convex problem. Simulation results illustrate that the significant performance improvement can be achieved by using the proposed practical model in the IRS-aided wideband OFDM system.

\section{Practical Model of IRS Reflection Coefficient for Wideband Signals}
The IRS is generally constructed as a printed circuit board (PCB). Particularly, each IRS reflecting element is composed of a metal patch on the top layer of the PCB and a full metal sheet on the bottom layer, which are connected to each other through a common varactor diode. The equivalent capacitance of each varactor changes according to the bias voltage, which can be adjusted by a controller. In other words, by tuning the equivalent capacitance, each IRS element is able to reflect the incident EM wave with different amplitude attenuations and phase shifts.

\begin{figure}[t]
\centering
  \vspace{-0.5 cm}
  \includegraphics[width= 3.5 in]{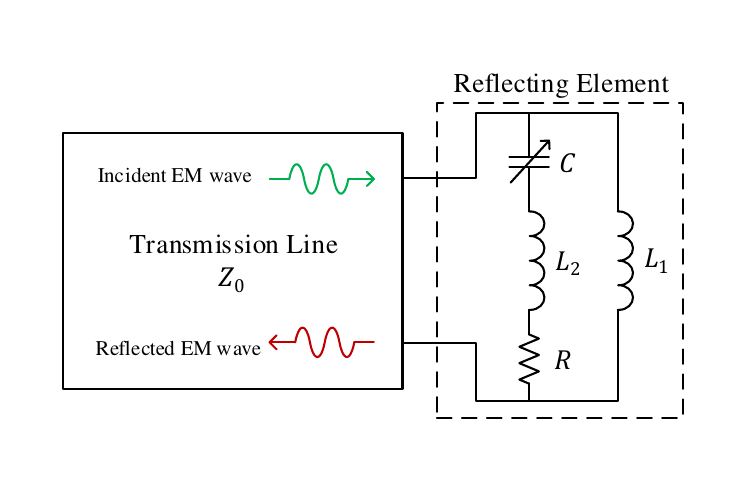}
  \vspace{-0.5 cm}
  \caption{The equivalent circuit of a practical IRS element.}\label{fig:electric_line}
   \vspace{-0.3 cm}
\end{figure}
As shown in Fig. \ref{fig:electric_line}, the electrical characteristics of an IRS element is illustrated by a parallel resonant circuit.
The amplitude and phase shift of the reflected signal are controlled by selecting appropriate capacitance $C$.
To be specific, for a signal of frequency $f$, the impedance of the reflect element circuit can be written by \vspace{-0.1 cm}
\begin{equation}
Z(C,f) = \frac{j2{\pi}f L_1(j2{\pi}f L_2+\frac{1}{j2{\pi}f C}+R)}{j2{\pi}fL_1+(j2{\pi}fL_2+\frac{1}{j2{\pi}f C}+R)} , \vspace{-0.1 cm}
\end{equation}
where $L_1$, $L_2$, $C$, and $R$ denote the bottom layer inductance, top layer inductance, equivalent variable capacitance and the loss resistance in the equivalent circuit, respectively.

Then, the reflection coefficient $\phi $ of the IRS element can be obtained by \vspace{-0.1 cm}
\begin{equation}
\phi(C, f) = \frac{{Z}({C},f)-{Z}_{0}}{{Z}({C},f)+{Z}_{0}},\label{eq:gamma} \vspace{-0.1 cm}
\end{equation}
which essentially describes the reflection of EM wave due to the impedance discontinuity between the free space impedance $Z_0 = 377 \Omega$ and the element parallel resonant circuit impedance $Z(C,f)$ \cite{R1}.
Obviously, the reflection coefficient $\phi$ is a function of $f$ with given/fixed electrical characteristics. This implies that the same IRS elements with determined circuit parameters generate different responses (i.e. reflection coefficients) to the signals with different frequencies.

\begin{figure}[!t]
\centering
  \vspace{-0.2 cm}
  \includegraphics[height= 3 in]{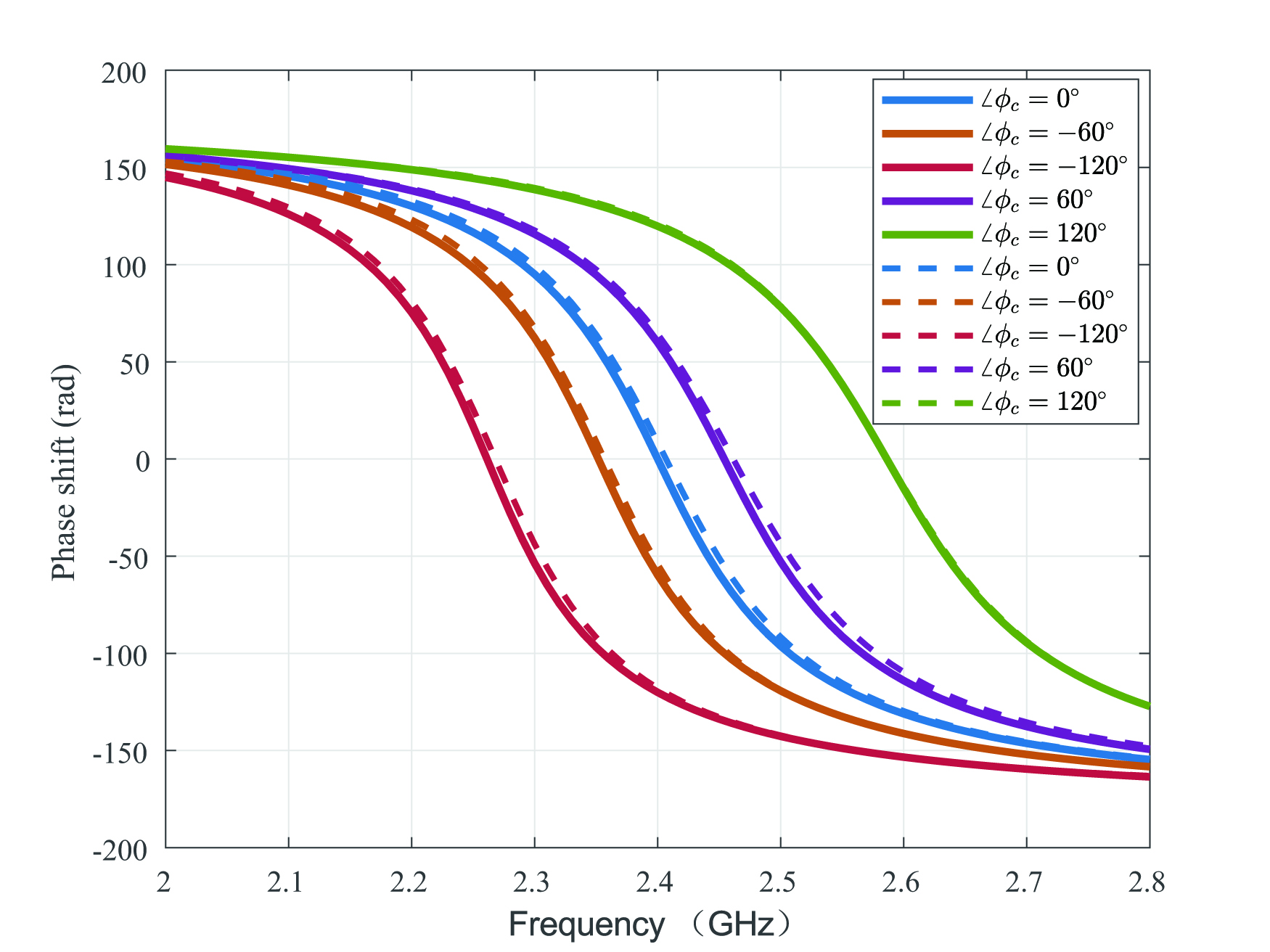}
  \vspace{-0.2 cm}
  \caption{The reflection phase shift varies with frequency. Solid: numerical results, dash: theoretical model.
  }\label{fig:angle}
 \vspace{0.3 cm}
  \includegraphics[height= 3 in]{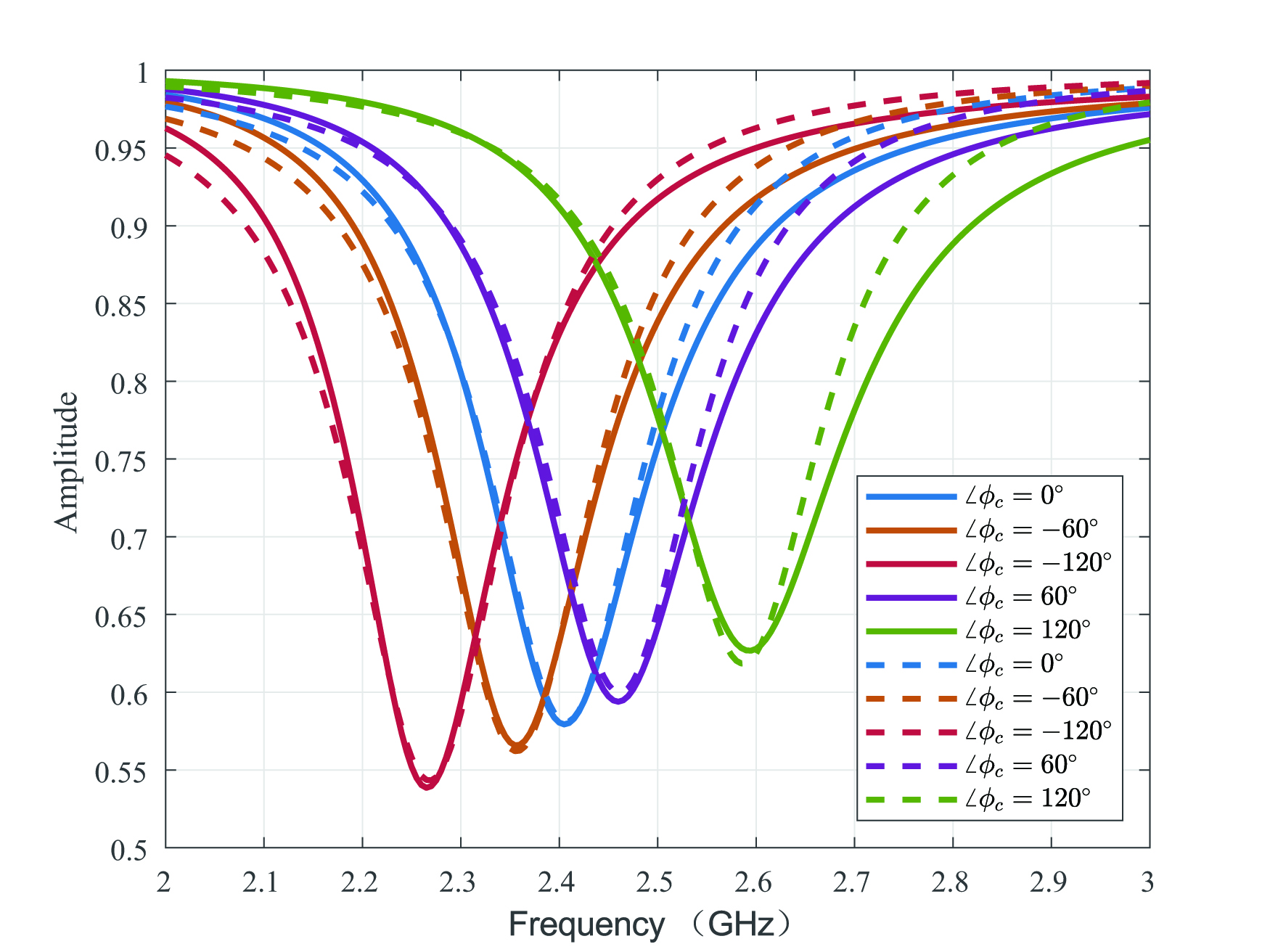}
  \vspace{-0.2 cm}
  \caption{The reflection amplitude varies with frequency. Solid: numerical results, dash: theoretical model.}\label{fig:abs}
  \vspace{-0.3 cm}
\end{figure}

To illustrate the importance of this phenomenon, we consider a wideband system with center frequency at  $2.4 \mathrm{GHz}$.
To assist such system, the practical surface-mount diode in IRS has $L_2=0.7 \mathrm{nH}$, $R=1 \Omega$, and $C$ varying from $0.47 \mathrm{pF}$ to $2.35 \mathrm{pF}$, which according to the device manual for the diode SMV1231-079. We also assume $L_1=2.5 \mathrm{nH}$ as in \cite{R2}.
Let $\phi_\mathrm{c}$ denote the reflection coefficient for the signal at the center frequency.
By selecting a appropriate value of $C$, the phase shift of $\phi_\mathrm{c}$  is adjusted to $\angle\phi_\mathrm{c}=0^{\circ}$. Then, the phase shifts for signals with different frequencies are calculated and plotted in Fig. \ref{fig:angle} as the solid blue curve.
It can be observed that, when the IRS element can provide $0^{\circ}$ phase shift for 2.4GHz signal, it will produce quite different phase shifts for the signals with different frequencies.
For example, the phase shift can reach $-100^{\circ}$ for 2.5GHz signal.
We repeat the simulation with $\angle\phi_\mathrm{c}= \pm 60^{\circ}, \pm 120^{\circ}$, respectively, and the similar conclusion can be found in Fig. \ref{fig:angle}.

In a similar way, we can also calculate the amplitudes of the reflected signals with different frequencies and plot the amplitudes as a function of frequency in Fig. \ref{fig:abs}, where different curves represent different phase shifts at the center frequency.
It is worth noting that the amplitudes of the reflected signals heavily depend on the frequency. The minimum amplitude appears when the frequency of the incident signal is the same as the resonant frequency of the reflection circuit. This is because the reflective current is in-phase with the element voltage in such a situation, which leads to the highest dielectric loss, metallic loss, and ohmic loss.

In an effort to characterize this fundamental relationship between phase shift/amplitude and frequency for optimizing IRS-aided wideband systems, we propose an analytical model to easily describe the phase shift and amplitude for the signals with different frequencies. We define the reflection coefficient of the $n$-th element of IRS as
\begin{equation}
\phi_n =  A_n e^{j\theta_n}, \forall n, \label{eq:gamma1}
\end{equation}
where $A_n \in [0, 1]$ and $\theta_n \in [-\pi, \pi)$ are the amplitude and phase of $\phi_n$, respectively.

Based on the shapes of curves in Figs. \ref{fig:angle} and \ref{fig:abs}, the phase shift variations are close to the horizontally-flipped arc-tangent function and the amplitude variations can be approximated as a vertically-flipped \textit{Witch of Agnesi} curve. Therefore, the phase shift $\theta_n(\angle{\phi_\mathrm{c}},f)$ and amplitude $A_n(\angle{\phi_\mathrm{c}},f)$ can be modeled as functions of $\angle{\phi_\mathrm{c}}$ and $f$ as
\begin{equation}
\begin{aligned}
 \theta_n(\angle{\phi_\mathrm{c}},f) &= -2\tan^{-1}[\mathcal{F}_2(\angle{\phi_\mathrm{c}})(f/{10^9}-\mathcal{F}_1(\angle{\phi_\mathrm{c}}))], \forall n,   \\
A_n(\angle{\phi_\mathrm{c}},f) &= -\frac{\alpha_4\angle{\phi_\mathrm{c}}+\beta_3}{((f/{10^9}-\mathcal{F}_1(\angle{\phi_\mathrm{c}}))/0.05)^2+4}+1 , \forall n, \\
\mathcal{F}_1(\angle{\phi_\mathrm{c}}) &= \alpha_1\tan(\angle{\phi_\mathrm{c}}/3)+\alpha_2\sin(\angle{\phi_\mathrm{c}})+\beta_1,\\
\mathcal{F}_2(\angle{\phi_\mathrm{c}}) &= \alpha_3\angle{\phi_\mathrm{c}} + \beta_2,
\end{aligned}
\end{equation}
where $\alpha_1, \alpha_2, \alpha_3, \alpha_4, \beta_1, \beta_2, \beta_3$ are the parameters.
The values of parameters are related to the specific circuit implementation and can be easily obtained by a standard curve fitting tool.
For practical examples as shown in Figs. \ref{fig:angle} and \ref{fig:abs}, the appropriate parameters are summarized in Table \ref{table:1}. As illustrated in Figs. \ref{fig:angle} and \ref{fig:abs}, the proposed practical phase shift and amplitude models (dash curves) closely match the numerical results (solid curves). In the next section, we intend to apply this practical model for optimizing an IRS-aided wideband OFDM communication system to illustrate the importance of the practical model and validate the effectiveness of our proposed practical model.

\begin{table}[t]
\caption{The parameters of the proposed model.}
\vspace{-0.3 cm}
\begin{center}
\begin{tabular}{ccccccc}
\hline
$\alpha_1$ & $\alpha_2$ & $\alpha_3$ & $\alpha_4$ & $\beta_1$ & $\beta_2$ & $\beta_3$\\
\hline
0.2 & -0.015 & -0.75  & -0.05 & 2.4 & 11.02 & 1.65\\
\hline\label{table:1} \vspace{-1.2 cm}
\end{tabular}
\end{center}
\end{table}

\vspace{-0.2 cm}

\section{Joint Power allocation and Reflecting Design for Wideband OFDM Systems}
\begin{figure}[t]
\centering
\vspace{-0.5 cm}
  \includegraphics[height= 3 in]{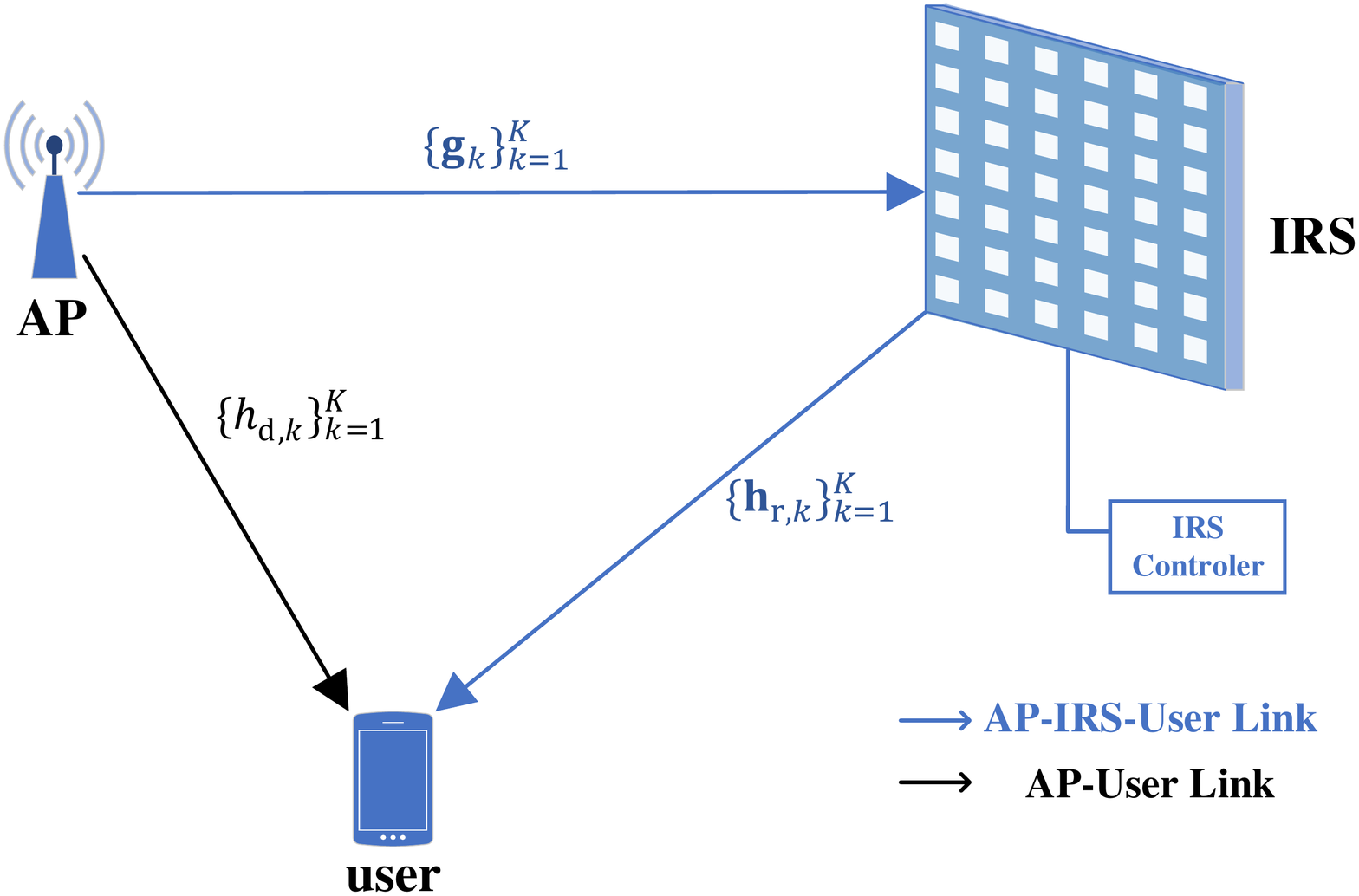}
  \vspace{-0.2 cm}
  \caption{An IRS-aided OFDM communication system.}\label{fig:System model}
  \vspace{-0.3 cm}
\end{figure}

We consider an OFDM system with $K$ subcarriers, as shown in Fig. \ref{fig:System model}, where an IRS composed of $N$ reflecting elements is utilized to assist the communication from a single-antenna access point (AP) to a single-antenna user. Let $ h_{\mathrm{d},k} \in \mathbb{C}$, $ \mathbf{h}_{\mathrm{r},k} \in \mathbb{C}^N$,
and $ \mathbf{g}_{k} \in \mathbb{C}^N$,
denote the baseband equivalent channels of the $k$-th subcarrier, $k = 1, \ldots, K$, from the AP to the user, from the IRS to the user, and from the AP to the IRS, respectively. We assume that channel state information (CSI) is known perfectly\footnote{The CSIs of IRS channels can be acquired by AP with effective channel estimation algorithms \cite{Online 1}, and the channel remains approximately constant within the transmission frame of our interest \cite{R3}.} and instantaneously to the AP, which can find the optimal IRS reflection and control the IRS via a dedicated control channel.

Let $ \angle{\phi_{\mathrm{c},n}}$, $n=1,\ldots, N$, be the phase shift of the $n$-th IRS element for the center frequency signal,
which is controlled by tuning the bias voltage (i.e. the capacitance) of the varactor.
In realistic applications, the phase shift of each IRS element at center frequency is controlled by $B$ bits and $ \angle{\phi_{\mathrm{c},n}}$ has $2^B$ discrete phase values\footnote{This work assumes uniformly quantizing the interval $ [-\pi,\pi)$ \cite{MY2}. Non-uniform quantization will be investigated in the future studies.}, i.e.
\begin{equation}
\angle{\phi_{\mathrm{c},n}} \in \mathcal{S} \triangleq \{\frac{2\pi}{2^B}b-\pi \ | \ b = 0, \ldots, 2^B-1\}, \forall n.
\end{equation}
Let $\phi_{n,k}$, $n=1,\ldots, N$, $k=1,\ldots, K$, denote the reflection coefficient of the $n$-th IRS element for the $k$-th subcarrier.
Then, with a phase value $ \angle{\phi_{\mathrm{c},n}}$ of a given reflect circuit, $\phi_{n,k}$ can be calculated by the proposed practical model in (4).

The received baseband signal for the $k$-th subcarrier at the user can be expressed as
\begin{equation}
y_k = \sqrt{p_k}(\mathbf{h}_{\mathrm{r},k}^H\mathbf{\Phi}_k\mathbf{g}_k+h_{\mathrm{d},k})s_k+n_k,\\
\forall k,\label{eq:receive2}
\end{equation}
where $p_k$ represents the signal transmission power for the $k$-th subcarrier, $s_k$ denotes the transmit symbol of each subcarrier and $n_k$ is the additive white Gaussian noise (AWGN) at the receiver with zero mean and variance $\sigma^2$, $\mathbf{\Phi}_k = \mathrm{diag}(\phi_{1,k}, \ldots, \phi_{N,k})$ is the reflect beamforming for the $k$-th subcarrier \footnote{The signals reflected by IRS more than once are ignored due to significant attenuation \cite{HY2}.}. Thus, the average rate over all subcarriers of this OFDM system can be calculated as
\vspace{-0.1 cm}
\begin{equation}
R = \frac{1}{K}\sum_{k=1}^K\mathrm{log}_\mathrm{2}\left(1+\frac{p_k|\mathbf{h}_{\mathrm{r},k}^H\mathbf{\Phi}_k\mathbf{g}_k+h_{\mathrm{d},k}|^2}{\sigma^2}\right) . \label{eq:rse}
\end{equation}

\vspace{-0.1 cm}
In this letter, we aim to jointly optimize the AP power allocation $\mathbf{p} \triangleq [p_1,\ldots,p_K]^T$ and the IRS reflect beamforming $\bm{\phi}_\mathrm{c} = [\angle{\phi_{\mathrm{c},1}, \ldots, \phi_{\mathrm{c},N}}]^T$ to maximize the average rate $R$, which is formulated as:
\vspace{-0.1 cm}
\begin{equation}
\begin{aligned}
\max\limits_{\mathbf{p},\bm{\phi}_\mathrm{c}}\ &R \\
\textrm{s.t.} \,\,\,\,  &\sum\nolimits_{k=1}^{K}p_k \leq P,\\
&\phi_{k,n} = A_n(\angle{\phi_{\mathrm{c},n}}, f_k)e^{j\theta_n(\angle{\phi_{\mathrm{c},n}},f_k)},\forall {k,n},\\
&\angle{\phi_{\mathrm{c},n}}\in\mathcal{S}, \forall n, \\
\end{aligned} \label{eq: P1}
 \end{equation}
where $P$ denotes the maximum transmit power at the AP. In order to efficiently solve this non-convex problem, we propose to iteratively find the conditionally optimal solution of $\bm{\phi}_\mathrm{c}$ and $\mathbf{p}$.
With given $\mathbf{p}$, problem (\ref{eq: P1}) can be reformulated as
\vspace{-0.1 cm}
\begin{equation}
\begin{aligned}
\max\limits_{\bm{\phi}_\mathrm{c}} & \;\;\; R
\\
\textrm{s.t.}  \,\,\,\, &\phi_{k,n} = A_n(\angle{\phi_{\mathrm{c},n}}, f_k)e^{j\theta_n(\angle{\phi_{\mathrm{c},n}}, f_k)},\forall {k,n},\\
&\angle{\phi_{\mathrm{c},n}}\in \mathcal{S}, \forall n,
\end{aligned}
\end{equation}
which is still non-convex and difficult to solve. We propose to successively find the conditionally optimal phase shift at the center frequency for one element with given others. Particulary, to find the optimal $\angle{\phi_{\mathrm{c},n}}$, we examine all possible values from set $\mathcal{S}$ and calculate the corresponding reflection coefficient for other subcarriers/frequences based on the proposed model.
Then, the best one which can provide the largest average rate is selected for $\angle{\phi_{\mathrm{c},n}}$.
The proposed reflect beamforming design is summarized in Algorithm 1.
Once $\bm{\phi}_\mathrm{c}$ is determined, the conditionally optimal power allocation $\mathbf{p}$ can be easily obtained by the water-filling algorithm. Now, the complete procedure can be established by iteratively optimizing $\bm{\phi}_\mathrm{c}$ and $\mathbf{p}$ until the convergence is found.

\begin{algorithm}[!t]
\caption{Reflect beamforming design.}
\label{alg:SH}
    \begin{algorithmic}[1]
    \begin{small}
    \REQUIRE $\mathbf{h}_{\mathrm{r},k}, \mathbf{g}_k, h_{\mathrm{d},k}, \mathcal{S}, \sigma$.
    \ENSURE $\bm{\phi}_{\mathrm{c}}^{*}$.
        \STATE {Initialize $ \angle{\phi_{\mathrm{c},n}} $, $n = 1, \ldots, N$. }
        \FOR {$n=1$ to $N$}
        \FOR  {$q=1$ to $|\mathcal{S}|$}
        \STATE Assign a value to $\angle{\phi_{\mathrm{c},n}} \in \mathcal{S}$.
        \FOR {$k=1$ to $K$}
            \STATE {Let $\phi_{k,n} = A_n(\angle{\phi_{\mathrm{c},n}}, f_k)e^{j\theta_n(\angle{\phi_{\mathrm{c},n}},f_k)}$.}
        \ENDFOR
            \STATE {Calculate $R$ by (\ref{eq:rse}).}
        \ENDFOR
            \STATE {Find the $\angle{\phi^{*}_{\mathrm{c},n}}$ which has maximum $R$.}
        \ENDFOR
        \STATE {Goto step 2 while no convergence of $R$. }
        \STATE {$\bm{\phi}_\mathrm{c}^{*} = [\angle{\phi^{*}_{\mathrm{c},1}},  \ldots, \angle{\phi^{*}_{\mathrm{c},N}}]$. }
        \end{small}\vspace{-0.0 cm}
    \end{algorithmic}
\end{algorithm}
\vspace{-0.3 cm}

\section{Simulation Results}
\begin{figure}[!t]
\centering
\vspace{-0.3 cm}
\includegraphics[height=3 in]{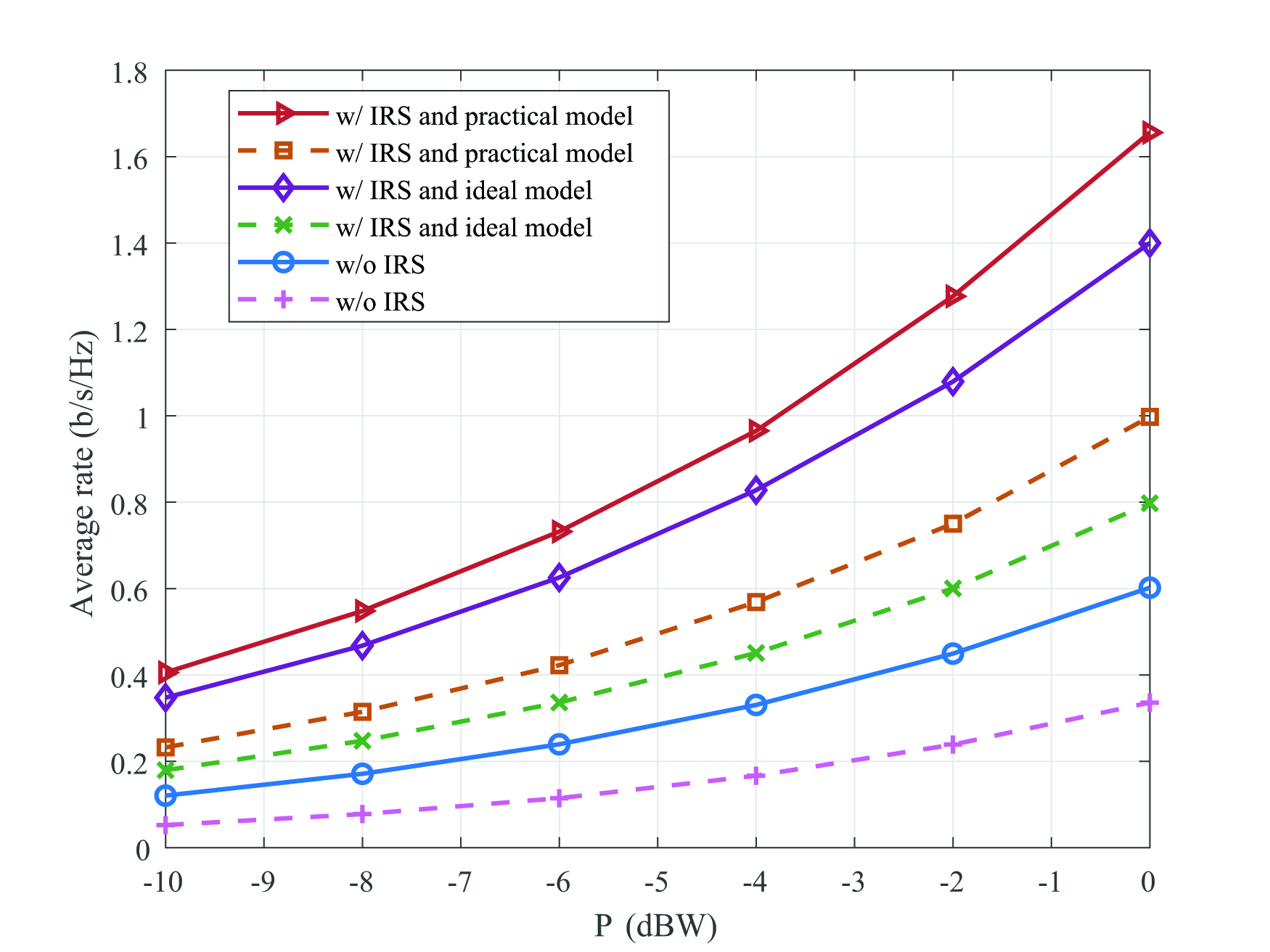}\vspace{-0.2 cm}
\caption{\small Average rate versus transmit power $P$. Solid: 100MHz bandwidth; dash: 200MHz bandwidth.}\label{fig:result}
\includegraphics[height=3 in]{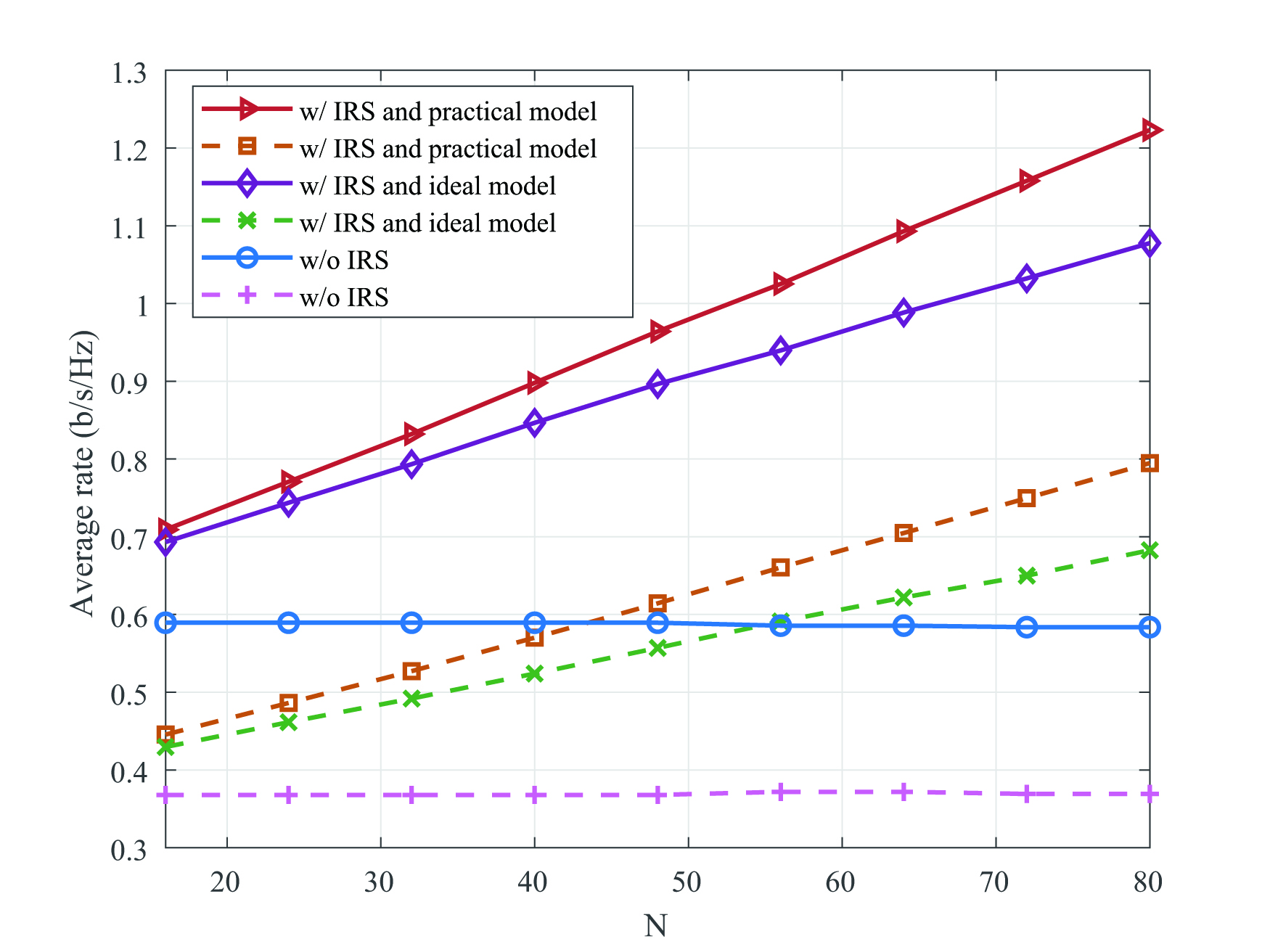}\vspace{-0.2 cm}
\caption{\small Average rate versus number of IRS elements $N$. Solid: 100MHz bandwidth; dash: 200MHz bandwidth.}\label{fig:result2}
\vspace{-0.4 cm}
\end{figure}

We consider an OFDM system consisting of a single-antenna the AP, a single-antenna user and an IRS with $N = 128$ reflecting elements to aid the communication.
Both IRS and user are located 50 meters (m) apart from AP.
Moreover, the user is randomly located around the IRS with 2m distance between them.
We assumed that all the channels involve rayleigh fading and the signal attenuation at a reference distance of 1m is set as $30 \mathrm{dB}$. The path loss exponents are set to 2.5, 2.8, and 3.5 for the channels between AP-IRS, IRS-user, and AP-user, respectively. In our simulation, we assume that IRS has eight possible discrete phase shifts controlled by 3 bits.
Fig. \ref{fig:result} shows the average achievable rates by using the proposed model and the ideal model for optimizing the IRS system. Two OFDM systems are evaluated: \textit{i}) $100 \mathrm{MHz}$ bandwidth with $K = 64 $ subcarriers (solid curves) and \textit{ii}) $200 \mathrm{MHz}$ bandwidth with $K = 128 $ subcarriers (dash curves). In the simulation results, the curves with the legend ``w/IRS and ideal model'' show the sum-rate achieved when employing  ideal model during beamforming design, while the proposed realistic model is used to compute performance. The curves with the legend ``w/IRS and practical model'' are obtained by using the practical model for both design and sum-rate calculation. In addition, the performance without using the IRS (w/o IRS) is also included as a benchmark.
It is observed from Fig. \ref{fig:result} that, while the IRS can significantly improve performance, our proposed practical model can provide remarkable better performance than the ideal model for optimizing the wideband IRS-aided OFDM systems. Interestedly, the system with wider bandwidth has less performance improvement. This is because that when frequency space is fixed, the more subcarriers, the more phase-amplitude-frequency distortion will be produced, which causes difficulty on the reflect beamforming optimization.
Moreover, Fig. \ref{fig:result2} illustrates the performance of different schemes as a function of the number of IRS elements.
In addition to the similar conclusion as in Fig. \ref{fig:result}, it is worth noting that performance improvement is also proportional to the number of IRS elements.

\vspace{-0.3 cm}
\section{Conclusions}
In this letter, we illustrated that the reflection of the incident signal by an IRS is heavily associated with the signal frequency and then proposed a practical model of the reflection coefficient for wideband IRS-aided systems. Based on this practical model, joint power allocation and reflecting optimization algorithm was introduced for an IRS-aided wideband OFDM system. Simulation results demonstrated the significant performance improvement by utilizing the proposed practical model in optimizing the IRS-aided system. While in this initial work a heuristic IRS design algorithm was developed to illustrate the importance of the practical model, possible directions for future studies include the design of more efficient algorithms and expand communication scenario.

\end{document}